\documentclass{article}
\usepackage{amsmath}

\usepackage{PRIMEarxiv}
\usepackage[utf8]{inputenc} 
\usepackage[T1]{fontenc}    
\usepackage{hyperref}       
\usepackage{url}            
\usepackage{booktabs}       
\usepackage{amsfonts}       
\usepackage{nicefrac}       
\usepackage{soul}
\usepackage{microtype}      
\usepackage{lipsum}
\usepackage{fancyhdr}       
\usepackage{graphicx}       
\graphicspath{{media/}}     
\usepackage{subfigure}
\usepackage{graphicx} 
\usepackage{tabularx} 
\usepackage{amsmath} 
\title{Neural network based model predictive control of voltage for a polymer electrolyte fuel cell system with constraints}

\author{
\textbf{Xiufei Li} $^{1\ast}$ \quad
\textbf{Miao Yang} $^{3\ast}$  \quad  
\textbf{Yuanxin Qi}  $^{1}$ \quad 
\textbf{Miao Zhang}  $^{2\ast\ast}$ \\
$^{1}$ Lund University, Department of Energy Sciences, Lund, 22100, Sweden  \\
$^{2}$ Shenzhen International Graduate School,Tsinghua University, Shenzhen, 518055, China\\
$^{3}$ City University of Hong Kong, Department of Mechanical Engineering, Hong Kong, 999077, China\\
$^{\ast\ast}$ Corresponding author: Miao Zhang, zhangmiao@sz.tsinghua.edu.cn\\
$^{\ast}$ First Author and Second Author contribute equally to this work.}

\begin{document}
\maketitle
\begin{abstract}
A fuel cell system must output a steady voltage as a power source in practical use.
A neural network (NN) based model predictive control (MPC) approach is developed in this work to regulate the fuel cell output voltage with safety constraints. The developed NN MPC controller stabilizes the polymer electrolyte fuel cell system's output voltage by controlling the hydrogen and air flow rates at the same time. The safety constraints regarding the hydrogen pressure limit and input change rate limit are considered. The neural network model is built to describe the system voltage and hydrogen pressure behavior. Simulation results show that the NN MPC can control the voltage at the desired value while satisfying the safety constraints under workload disturbance. The NN MPC shows a comparable performance of the MPC based on the detailed underlying system physical model.

\end{abstract}


\section{Introduction}
\label{Introduction}
The polymer electrolyte fuel cells (PEFCs) have been regarded as a promising power source due to their environmental friendliness, high energy density, and low operating temperature \cite{baroutaji2021advancements,zhang2020thermal,dong2021honeycomb,xu2021analysis} in contrast to the unpleasant environmental impacts from traditional fossil fuel usage. The PEFCs are more and more favored in stationary, portable and transport applications. However, its commercialization is still limited by some technical challenges, among which the reliable operation is an important topic \cite{liang2021electrocatalytic,cao2020efficient,parbey2020electrospun}. Control algorithms play a significant role in ensuring the PEFC systems' reliability. The high complexity of PEFC systems as well as the safety limits require more advanced control methods \cite{silaa2020design}. For the PEFC control application, MPC is a promising method because of its ability of handling the multi inputs and constraints. MPC has been widely used in PEFC systems for different applications. \cite{quan2021feedback} 
adopted a multi-input multi-output MPC approach for the control of hydrogen excess ratio and the balance of electrodes pressure.

 \cite{ghanaatian2018control} proposed an optimal nonlinear MPC controller for a flywheel energy storage system with constraints on the system states and actuators being taken into account. The simulation demonstrated that the proposed method was effective at controlling the DC-link voltage and the flywheel speed.  \cite{wang2013real} studied the modeling and air flow control for PEFCs using MPC. It was found that the proposed MPC presented superior performance as compared to the proportional-integral control. \cite{vahidi2004model} formulated a distribution of current demand between the fuel cell and the auxiliary source using an MPC framework to avoid stack starvation and damage. The results showed that the reactant deficit during sudden increases in stack power was reduced from 50\% in stand-alone architecture to less than 1\% in the hybrid configuration. 
A lot of research approaches to MPC applications can be found in recent studies and MPC was demonstrated to be an effective method.

However, the models used in the research mentioned above are mainly physical models. Many partial differential equations are involved in physical modeling. Building the model as well as its simplification requires human expert knowledge, making this process time-consuming. Alternatively, data-based modeling exploits the observed data instead of the inner physical laws to obtain the system model. Data-based models have attracted more and more attention because of their powerful representation capability, flexibility and easy-to-build properties. The data-based methods and control approaches based on those models are attractive to fuel cell applications due to the nonlinearity and complexity of fuel cell systems. \cite{MEHRPOOYA20188} first experimentally collected data with different inlet humidity, temperature, and oxygen and hydrogen ﬂow rates, then trained a two-hidden-layer neural network to predict fuel cell steady-state performance. To describe the fuel cell polarization curve, \cite{HAN201610202} adopted both neural network and support vector machine and compared their performances. \cite{BICER20161205} collected training data from MATLAB simulation, and trained neural network for fuel cell dynamics prediction.

There is also some research efforts focusing on the control methods using neural networks for fuel cell applications.
\cite{ELSHARKH200488} adopted a active and reactive power controller based on neural networks to manage a stand-alone PEFC output power.
\cite{HAJIMOLANA2013320} developed neural network predictive control for thermal stress management of a solid oxide fuel cell and compared the performance with the PI controller.
A genetic algorithm neural network based predictive control method was adopted by \cite{WU2008232} to control the fuel cell stack terminal voltage as a proper constant.

A hierarchical predictive control based energy management strategy for fuel cell hybrid construction vehicles was presented by \cite{LI2020117327}, where the neural network was used for the complex load prediction.

When used as a power generation source, a fuel-cell system should be able to provide a stable constant voltage \cite{li2021adaptive}. In our previous paper \cite{li2021multi}, an MPC implementation was developed to control the PEFC system's output voltage at a steady state by regulating its input hydrogen and air volumetric flow rates. In this paper, the safety constraints regarding hydrogen pressure limit and input change rate limit are added. Furthermore, a neural network MPC controller was developed to regulate the PEFC system's voltage under low current disturbances with safety constraints. The neural network is built to model the fuel cell voltage and hydrogen pressure. The performance of the NN MPC is shown and compared with the MPC based on the detailed system physical models.

The rest of the paper is organized as follows: Section 2 presents the neural network modeling process; Sec. 3 detailed the NN MPC algorithm development; Sec. 4 illustrated the simulation set-up; Sec. 5 shows the performance of the NN MPC and its comparison to MPC based on physical models; Sec. 7 exhibits the conclusion.

\section{Neural network modeling}
\label{Neural network modeling}

\subsection{Neural network}
\begin{figure}
    \includegraphics[width=0.8\linewidth]{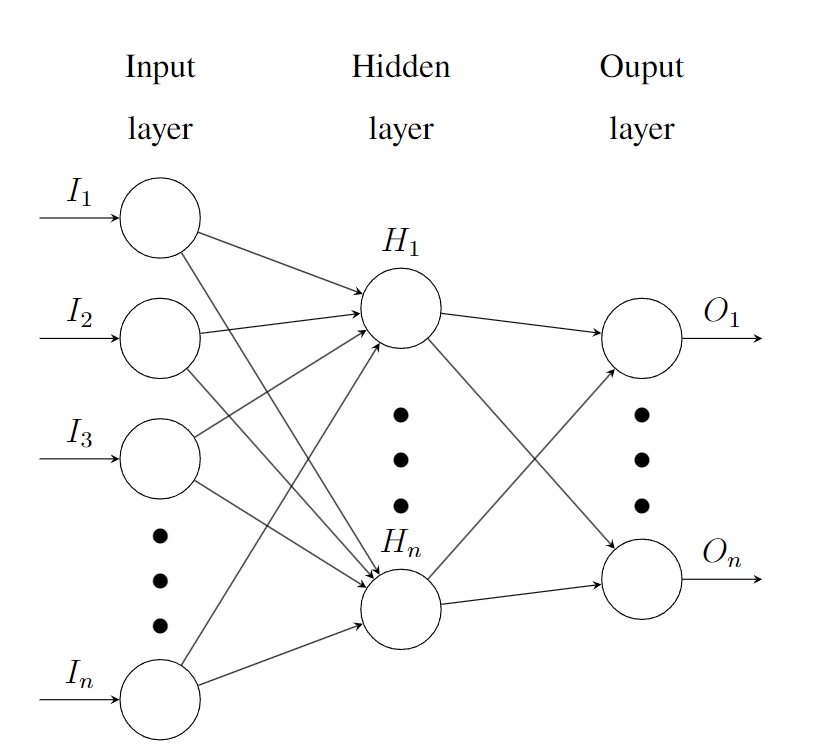}
    \centering
    \caption{Neural network structure}
    \label{fig:neuralnetworkstructure}
\end{figure}

A neural network is a connectionist system inspired by a biological neural network in the animal brain~\cite{anderson1995introduction}.
It makes great success in several topics and has been a hot research area in recent years due to the massive data available and cheap computational power.
Normally, it is constituted by the input layer, hidden layers, and output layer.
An illustration diagram of its structure can be found in Fig.~\ref{fig:neuralnetworkstructure}.
There can be more than one hidden layer in the network.
Each circle in Fig.~\ref{fig:neuralnetworkstructure} is a node, or a neuron, with an activation function defining the output with respect to the inputs after an affine transformation. Many kinds of activation functions are available. Each arrow shows a data flow.

For each hidden node, the output is:
\begin{equation} \label{eq:neuralnetwork}
\begin{array}{rl}
    \boldsymbol{h}(\boldsymbol{x})\!&\!= g(\boldsymbol{w}^T \boldsymbol{x} + \boldsymbol{b})  \\
     g(z)\!&\!= \max(0, z)
\end{array}
\end{equation}
where $\boldsymbol{h}$ and $\boldsymbol{x}$ are the unit outputs and inputs; $\boldsymbol{w}$ and $\boldsymbol{b}$ are the parameters to be optimized on; $g(z)$ is the activation function adopted in this work, which is called the rectified linear unit (ReLU).

\subsection{Neural network for fuel cell modeling}
The detailed physical fuel cell model can be found in our previous work \cite{li2021multi}. The model was elaborated and built with MATLAB Simulink, which was assumed to be the true system dynamics. In order to model errors in measurement, Gaussian measurement noises were added to output voltage $V_\mathrm{FC}$ and hydrogen pressure $P_\mathrm{H_2}$. 

The neural network model predicts two system states based on three system inputs. The inputs are the control actions, hydrogen volumetric flow rate $Q_\mathrm{H_2}$, air volumetric flow rate $Q_\mathrm{air}$, and the current $I$; the states are the values to be predicted, the fuel cell output voltage $V_\mathrm{FC}$ and the hydrogen pressure $P_\mathrm{H_2}$. It should be noticed that the current $I$ is the workload and can not be manipulated by the controller. The fuel cell voltage is the control target which should be kept at a fixed value. The hydrogen pressure needs to be kept below a certain limit to ensure safety.

The target of NN modeling is to build a function $\boldsymbol{f}$ that describes the fuel cell dynamics in the time update:
\begin{equation} \label{eq:2outputsNN}
    \boldsymbol{x}^{k+1} = \boldsymbol{f}(\boldsymbol{u}^k,\boldsymbol{x}^k)
\end{equation}

where
\begin{equation}
    \begin{array}{rcl}
        \boldsymbol{u}^k & = & [Q^k_\mathrm{H_2}\ Q^k_\mathrm{air}\ I^k]^T \\
        \boldsymbol{x}^k & = & [V^k_\mathrm{FC}\ P^k_\mathrm{H_2}]^T
    \end{array}
\end{equation}

the variable $\boldsymbol{u}^k$ is the model inputs at time step $k$; $\boldsymbol{x}^{k+1}$ is the system states at time step $k+1$ and $\boldsymbol{x}^k$ is the measured system states at time step $k$.

The network structure is (5, 16, 32, 8, 2), which means that the NN input layer has five nodes and the output layer has two nodes, and NN has three hidden layers with 16, 32, and 8 nodes respectively. To collect the training data points for the Gaussian process, Latin hypercube sampling (LHS) was applied to the inputs $Q_\mathrm{H_2}$, $Q_\mathrm{air}$ and the current $I$. The interval between time steps was 0.5 s. In total, 2000 training data points were collected.

\section{Neural network MPC design}
\label{NN MPC design}
The state-space model for MPC is obtained by linearizing the neural network model. The neural network $\boldsymbol{f}$ shown in Eq.~(\ref{eq:2outputsNN}) then becomes: 

\begin{equation}
    \begin{array}{rl}
        \mathrm{d}V_\mathrm{FC}&\!= \begin{bmatrix} \dfrac{\partial V^{k+1}_\mathrm{FC}}{\partial Q^{k}_\mathrm{H_2}} & \dfrac{\partial V^{k+1}_\mathrm{FC}}{\partial Q^{k}_\mathrm{air}} & \dfrac{\partial V^{k+1}_\mathrm{FC}}{\partial I^{k}}  \end{bmatrix} \! \begin{bmatrix} \mathrm{d}Q_\mathrm{H_2}\\ \mathrm{d}Q_\mathrm{air} \\ \mathrm{d}I \end{bmatrix} 
        \\
        \mathrm{d}P_\mathrm{H_2}&\!= \begin{bmatrix} \dfrac{\partial P^{k+1}_\mathrm{H_2}}{\partial Q^{k}_\mathrm{H_2}} & \dfrac{\partial P^{k+1}_\mathrm{H_2}}{\partial Q^{k}_\mathrm{air}} & \dfrac{\partial P^{k+1}_\mathrm{H_2}}{\partial I^{k}}  \end{bmatrix} \! \begin{bmatrix} \mathrm{d}Q_\mathrm{H_2}\\ \mathrm{d}Q_\mathrm{air} \\ \mathrm{d}I \end{bmatrix} 
    \end{array}
\end{equation}
where the partial derivative is taken on the latest system states, which will update each time step. The partial derivative of the neural network is solved by the forward mode automatic differentiation (AD) method \cite{baydin2018automatic}.

The discrete-time state-space model of the fuel cell used for control is written as:
\begin{equation} \label{eq:statespacemodelnnfc}
  \begin{array}{rl}
    \boldsymbol{x}^{k+1}&\!= \boldsymbol{A} \boldsymbol{x}^k + \boldsymbol{B} \boldsymbol{u}^k
    \\
    \boldsymbol{y}^k&\!= \boldsymbol{C} \boldsymbol{x}^k
  \end{array}
\end{equation}
where the state vector $\boldsymbol{x}^k$ at sample index $k$ is:
\begin{equation} \label{eq:ssmxnnfc}
  \boldsymbol{x}^k = \begin{bmatrix} V_\mathrm{FC}^{k}\\ P_\mathrm{H_2}^{k}\\ \mathrm{d}I^{k}\\ Q_\mathrm{H_2}^{k}\\ Q_\mathrm{air}^{k} \end{bmatrix}
\end{equation}
with input:
\begin{equation} \label{eq:ssmunnfc}
  \boldsymbol{u}^k = \begin{bmatrix} \mathrm{d}Q_\mathrm{H_2}^k\\ \mathrm{d}Q_\mathrm{air}^k  \end{bmatrix}
\end{equation}
and output:
\begin{equation} \label{eq:ssmynnfc}
  \boldsymbol{y}^k = \begin{bmatrix} V_\mathrm{FC}^{k} \end{bmatrix}
\end{equation}
and state-space matrices:
\begin{equation} \label{eq:ssmABCnnfc}
  \begin{array} {ll}
    \boldsymbol{A} & = \begin{bmatrix}
      1\ & 0\ & \frac{\partial V^{k+1}_\mathrm{FC}}{\partial I^{k}}\ & 0\ & 0\\
      0\ & 1\ & \frac{\partial P^{k+1}_\mathrm{H_2}}{\partial I^{k}}\ & 0\ & 0\\
      0\ & 0\ & 0\ & 0\ & 0\\
      0\ & 0\ & 0\ & 1\ & 0\\
      0\ & 0\ & 0\ & 0\ & 1\\
      \end{bmatrix} \\[2mm]
    \boldsymbol{B} & = \begin{bmatrix}
      \frac{\partial V^{k+1}_\mathrm{FC}}{\partial Q^{k}_\mathrm{H_2}}\ & \frac{\partial V^{k+1}_\mathrm{FC}}{\partial Q^{k}_\mathrm{air}} \\[2mm]
      \frac{\partial P^{k+1}_\mathrm{H_2}}{\partial Q^{k}_\mathrm{H_2}}\ & \frac{\partial P^{k+1}_\mathrm{H_2}}{\partial Q^{k}_\mathrm{air}} \\
      0\ & 0\\
      1\ & 0\\
      0\ & 1
      \end{bmatrix} \\ [14mm]
    \boldsymbol{C} & = \begin{bmatrix} 1\ 0\ 0\ 0\ 0\ \end{bmatrix}
  \end{array}
\end{equation}

The actuator increments were selected as the system inputs.
Consequently, $Q_\mathrm{H_2}^{k}$ and $Q_\mathrm{air}^{k}$ were added into the state vector to help impose appropriate constraints.

A Quadratic Programming (QP) problem will be solved at each time step to obtain {the optimal} control inputs:
\begin{equation} \label{eq:MPCcostfunctionnnfc}
  \min_{\boldsymbol{u}^0,\boldsymbol{u}^1,...,\boldsymbol{u}^{H_u-1}} \boldsymbol{J}(\boldsymbol{u}^k) = 
    \sum_{k=1}^{H_{p}} \left \| \boldsymbol{y}^k - \boldsymbol{r} \right \|_{\boldsymbol{Q}}^2
        +\sum_{k=0}^{H_{u}-1} \left \| \boldsymbol{u}^k \right \|_{\boldsymbol{R}}^2
        + \rho \sum_{k=1}^{H_{p}} \left \| \boldsymbol{\epsilon}^k \right \|^2
\end{equation}
subject to:
\begin{equation} \label{eq:MPCconstraintsnnfc}
  \begin{array}{rl}
    \boldsymbol{x}^{k+1} &\!= \boldsymbol{A^d} \boldsymbol{x}^k + \boldsymbol{B^d} \boldsymbol{u}^k\\
    \boldsymbol{y}^k &\!= \boldsymbol{C^d} \boldsymbol{x}^k\\
    \boldsymbol{u}_\mathrm{lb} &\!\leq \boldsymbol{u}^k \leq \boldsymbol{u}_\mathrm{ub}\\
    \mathrm{d}\boldsymbol{u}_\mathrm{lb} &\!\leq \boldsymbol{u}^{k} - \boldsymbol{u}^{k-1} \leq \mathrm{d}\boldsymbol{u}_\mathrm{ub}\\
    \boldsymbol{x}_\mathrm{lb} &\!\leq \boldsymbol{x}^k \leq \boldsymbol{x}_\mathrm{ub} + \boldsymbol{\epsilon}^k\\
    \boldsymbol{0} &\!\leq \boldsymbol{\epsilon}^k \\
    \boldsymbol{u}^{-1} &\!= \boldsymbol{u}_\mathrm{init}\\
    \boldsymbol{x}^0 &\!= \boldsymbol{x}_\mathrm{init}\\
    k &\!= 0, 1, \dots, H_p
  \end{array}
\end{equation}
where $H_p$ and $H_u$ are prediction and control horizon;
$k$ in the superscript represents the time step, and $k=0$ refers to the current time step;
$\boldsymbol{r}$ is the control reference;
$\boldsymbol{Q}$ and $\boldsymbol{R}$ are weight tuning parameters for reference tracking and control inputs;
$\boldsymbol{A^d}$, $\boldsymbol{B^d}$, and $\boldsymbol{C^d}$ are state-space matrices $\boldsymbol{A}$, $\boldsymbol{B}$ and $\boldsymbol{C}$ in discrete-time;
$\boldsymbol{u}_\mathrm{lb}$, $\boldsymbol{u}_\mathrm{ub}$, $\boldsymbol{x}_\mathrm{lb}$, and $\boldsymbol{x}_\mathrm{ub}$ are the lower bounds and upper bounds of inputs $\boldsymbol{u}$ and states $\boldsymbol{x}$; 
$\mathrm{d}\boldsymbol{u}_\mathrm{lb}$ and $\mathrm{d}\boldsymbol{u}_\mathrm{ub}$ are the lower bounds and upper bounds of inputs change rate;
$\boldsymbol{u}_\mathrm{init}$ is latest applied control inputs and $\boldsymbol{x}_\mathrm{init}$ is the latest measured value, the state feedback;
$\boldsymbol{\epsilon}$ is the slack variable introduced to soft constraints and $\rho$ is a nonnegative scalar to control the magnitude of penalizing soft constraint violations.
The slack variable $\boldsymbol{\epsilon}$ is to deal with the possibility of infeasibility by hard constraints and model imperfection. It is defined that their value is non-zero only if the constraints are violated. Then the violations are highly penalized in cost function by a large enough $\rho$ to let the optimizer have a strong incentive to keep $\boldsymbol{\epsilon}$ at zero.

In the fuel cell problem, the state constraint is the limitation of the hydrogen pressure $P_\mathrm{H_2}$. The hydrogen pressure in the pipe should be under a certain value to ensure safety, which is 2.5 atm here. Only slack variables for state upper bounds are introduced. The input $Q_\mathrm{H_2}$ is limited between 100 and 400 lpm (liters per minute) and $Q_\mathrm{air}$ is limited within 300 to 700 lpm. The change rate of the inputs is constrained within -40 to 20 lpm. The fuel cell system size is 6 kW.

The time step for the MPC controller is 0.5 s. The QP problem is solved at each step, then the solved $Q_\mathrm{H_2}$ + $\mathrm{d} Q_\mathrm{H_2}$ and $Q_\mathrm{air}$ + $\mathrm{d} Q_\mathrm{air}$ are applied to the fuel cell plant.

\section{Simulation set-up}\label{Experimental set-up}
To test the controller performance, a simulation was conducted on the Simulink model detailed in \cite{li2021multi}. This Simulink model is used as the underlying true system model for comparison. For the voltage $V_\mathrm{FC}$ and the hydrogen pressure $P_\mathrm{H_2}$ in the system outputs, Gaussian measurement noises were added.

\begin{figure}[htp]
    \centering
    \includegraphics[width=\linewidth]{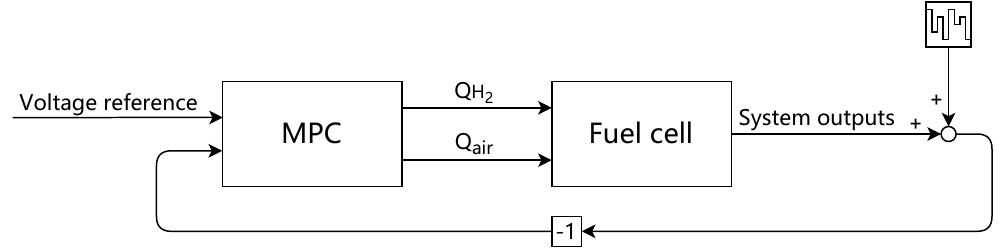}
    \caption{MPC control process}
    \label{fig:fuelcellcontroldiagram_MPC}
\end{figure}

Figure~\ref{fig:fuelcellcontroldiagram_MPC} shows the MPC control process.

The MPC formulation in Fig.~\ref{fig:fuelcellcontroldiagram_MPC} is detailed in \cite{li2021multi}. The constraints on the input change rate and hydrogen pressure were also added in this work.

\begin{figure}[htp]
    \centering
    \includegraphics[width=\linewidth]{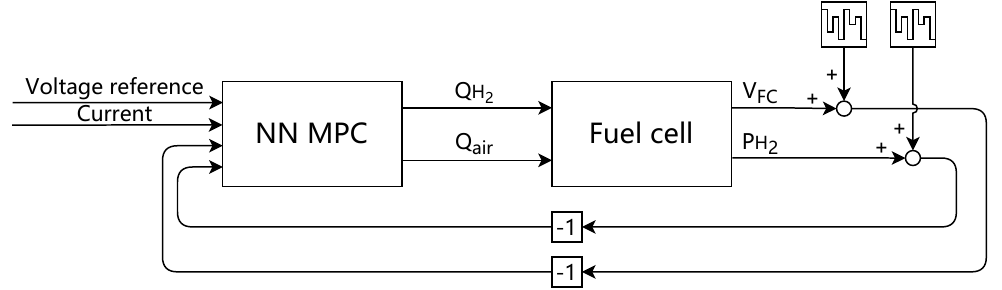}
    \caption{NN MPC control process}
    \label{fig:fuelcellcontroldiagram_NNMPC}
\end{figure}

Figure~\ref{fig:fuelcellcontroldiagram_NNMPC} illustrates the NN MPC control process.

In the NN MPC control process, only the system information of $V_\mathrm{FC}$ and $P_\mathrm{H_2}$ is needed. However, the MPC controller shown in Fig.~\ref{fig:fuelcellcontroldiagram_MPC} requires more system information, like the oxygen and nitrogen pressure. This is one advantage of the data-based modeling method used here.

\section{Simulation results}
\label{Simulation results}

The performance of NN MPC and MPC with physical models explained in \cite{li2021multi} were compared.
Two test scenarios were chosen, one was the typical step disturbance applied on the working load, the current; the other was a mixture of slope and step working load changes.

\begin{figure}[!bhtp]
    \centering
    \includegraphics[width=0.8\linewidth]{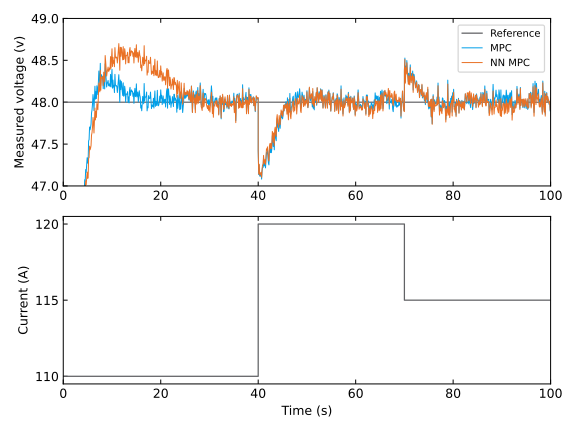}
    \caption{Output voltage under the current disturbance.}
    \label{fig:MPCvsnnMPC1}
\end{figure}

\begin{figure}[!bhtp]
    \centering
    \includegraphics[width=0.8\linewidth]{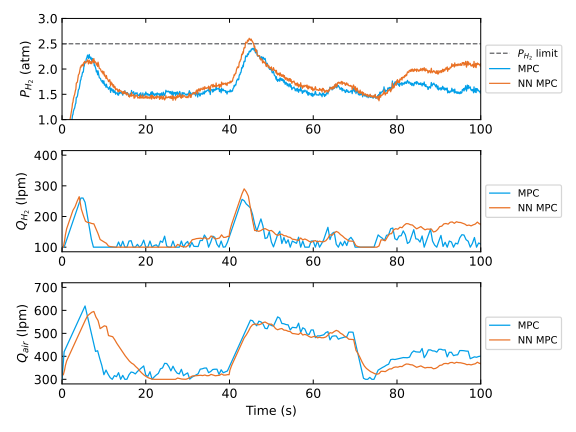}
    \caption{Constraint handling and system inputs.}
    \label{fig:MPCvsnnMPC2}
\end{figure}

Figure~\ref{fig:MPCvsnnMPC1} shows the NN MPC voltage tracking performance as compared to the MPC based on physical models under step workload disturbance, and Fig.~\ref{fig:MPCvsnnMPC2} gives the corresponding hydrogen pressure $P_\mathrm{H_2}$ behavior and system inputs. At the beginning of the simulation, the MPC and NN MPC had similar rise-up traces. This rise-up rate was limited by the input change rate constraint.
But later the MPC had a lower overshoot than NN MPC. The overshoot for MPC was 0.42 V and for NN MPC 0.57 V.
The MPC benefited from the knowledge of the true system physical model.
Both controllers can successfully satisfy the hydrogen pressure $P_\mathrm{H_2}$ safety requirements in this period. When the current suddenly increased, the MPC and NN MPC drove the voltage back to the reference with the same rate. It can be clearly seen that the input increment followed a straight line, indicating the activation of the input change rate limits.
But the MPC was able to suppress the hydrogen pressure under the safety limit, whereas NN MPC violated the constraint with a small peak in this step current increase scenario. The higher overshoot and the small constraint violation were due to the NN model imperfection and the error introduced by linearization. The safety constraint was satisfied throughout the remaining process.
When the current suddenly dropped at time 75 s, MPC and NN MPC had a similar performance in stabilizing the voltage at 48 V. The steady-state tracking behavior of the two controllers was comparable.


\begin{figure}
    \centering
    \includegraphics[width=0.8\linewidth]{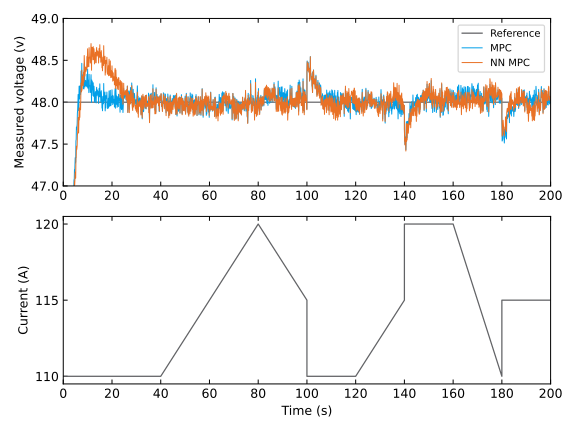}
    \caption{Output voltage under the current disturbance.}
    \label{fig:MPCvsNNMPC_rampI1}
\end{figure}

\begin{figure} 
    \centering
    \includegraphics[width=0.8\linewidth]{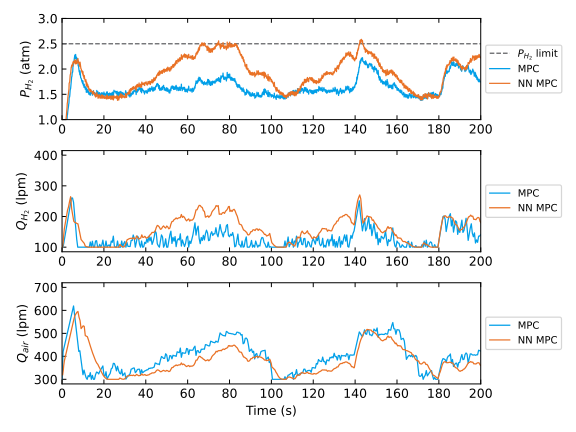}
    \caption{Constraint handling and system inputs.}
    \label{fig:MPCvsNNMPC_rampI2}
\end{figure}

Figures~\ref{fig:MPCvsNNMPC_rampI1} and~\ref{fig:MPCvsNNMPC_rampI2} present the NN MPC and MPC behavior under a mixture of slope and step current changes. The behavior was similar to the step change scenario. The NN MPC had a higher overshoot after the system start-up, and NN MPC violated the hydrogen pressure constraint with a small peak in this step current increase scenario at 140 s. In the period around 65 - 85 s, NN MPC kept the hydrogen pressure under the 2.5 atm limit with the effort of adjusting the input hydrogen flow rate, while the MPC had better handling by applying a lower hydrogen flow rate and a higher air flow rate in advance.

\section{Conclusion}

In summary, a neural network MPC was developed to control the fuel cell voltage under current disturbance with safety constraints. The neural network was built to describe the voltage and hydrogen pressure behavior. The state-space models in MPC design were formed based on neural network linearization. The simulation results showed that the proposed NN MPC method was able to stabilize the output voltage at the reference value under workload disturbance, and had a comparative performance of MPC controller with the knowledge of underlying system dynamics. The NN MPC could satisfy the hydrogen pressure limit in most cases, but it violated the constraint with a small peak in some step current increase scenarios.

\section*{Acknowledgements}
The first author would like to acknowledge the Competence Centre Combustion Processes, KCFP, and the Swedish Energy Agency (grant number 22485-4) for the financial support.
The Chinese Scholarship Council is also thanked for the sponsorship of living expenses during the first author’s research.
Rolf Johansson is a member of the LCCC Linnaeus Center and the eLLIIT Excellence Center at Lund University.


\bibliography{mybib.bib} 
\bibliographystyle{IEEEtran}
\end{document}